\documentclass[aip, reprint]{revtex4-1} 
\usepackage{amsmath,amssymb,amsfonts, graphicx,color,multirow}

\newcommand{\hbe}{\mathbf{\hat{e}}}

\newcommand{\p}{\partial}
\newcommand{\rmd}{\mathrm{d}}

\newcommand{\bv}{\mathbf{v}}

\newcommand{\bj}{\mathbf{j}}
\newcommand{\bs}{\mathbf{\sigma}}
\newcommand{\bU}{\mathbf{U}}

\newcommand{\tbv}{\tilde{\mathbf{v}}}

\newcommand{\tphi}{\tilde{\phi}}
\newcommand{\tc}{\tilde{c}}
\newcommand{\tr}{\tilde{r}}
\newcommand{\tU}{\tilde{U}}

\newcommand{\talpha}{\tilde{\alpha}}
\newcommand{\hc}{\hat{c}}

\newcommand{\hm}{\hat{m}}
\newcommand{\hM}{\hat{M}}
\newcommand{\hpsi}{\hat{\psi}}

\newcommand{\calR}{\mathcal{R}}
\newcommand{\calC}{\mathcal{C}}
\newcommand{\calU}{\mathcal{U}}
\newcommand{\calD}{\mathcal{D}}
\newcommand{\calNa}{\mathcal{N}_{A}}
\newcommand{\calkT}{\mathit{k}_{b} \mathcal{T}}
\newcommand{\ite}{\mathit{e}}
\newcommand{\iteta}{\mathit{\mathbf{\eta}}}
\newcommand{\iteps}{\mathit{\mathbf{\epsilon}}}

\newcommand{\Pe}{\mathrm{Pe}\,}

\begin{document}
\title{Nonlinear, electrocatalytic swimming in the presence of salt}
\author{Benedikt \surname{Sabass}}
\author{Udo \surname{Seifert}} 
\affiliation{II. Institut f\"ur Theoretische Physik, Universit\"at Stuttgart,
70550 Stuttgart, Germany}

\begin{abstract}
A small, bimetallic particle in a hydrogen peroxide solution can propel itself by
means of an electrocatalytic reaction. The swimming is driven by a flux of ions
around the particle. We model this process for the presence of a monovalent salt, where 
reaction-driven proton currents induce salt ion currents. A theory for thin diffuse layers is employed, which yields nonlinear, coupled transport equations.
The boundary conditions include a compact Stern layer of adsorbed ions. Electrochemical processes on the particle surface are modeled with a first order reaction
of the Butler-Volmer type.  The equations are solved numerically for the swimming speed. An analytical approximation is derived under the assumption that the decomposition of hydrogen peroxide occurs 
mainly without inducing an electric current. We find that the swimming speed increases linearly with hydrogen peroxide concentration for small concentrations.
The influence of ion diffusion on the reaction rate can lead to a concave shape of the function of speed \textsl{vs.}
hydrogen peroxide concentration. The compact layer of ions on the particle diminishes the reaction rate and consequently reduces the 
speed. Our results are consistent with published experimental data.
\end{abstract}
\maketitle

\section{Introduction}
Motion of micrometer-sized swimmers, directly driven by an inhomogeneous
chemical
surface reaction, has recently attracted much scientific interest. 
While the mechanism was originally suggested as a mode of biological self
propulsion\cite{mitchell1956hypothetical,lammert1996ion}, 
 currently studied swimmers are mostly artificial.
A variety of systems featuring this chemical self-propulsion, have been
investigated\cite{browne2006making, ebbens2010pursuit}. 
The employed reactions often involve hydrogen peroxide which decomposes on the
metal
\cite{imagilov2001,paxton2005motility,fournier2005synthetic, howse2007self,
erbe2008various}
or enzyme-coated\cite{vicario2005catalytic} surfaces of the swimmers. Other
mechanisms are, e.g., oxidation and reduction of
glucose\cite{mano2005bioelectrochemical},
hydrolyzation\cite{hanczyc2007fatty} or bromination\cite{thutupalli2011simple} at
droplet surfaces,
and thermally induced phase separation around a
swimmer\cite{volpe2011microswimmers}. Finally, a number of theoretical studies
 have been conducted with generic chemical reaction schemes
\cite{golestanian2005propulsion, ruckner2007chemically, thakur2011dynamics}.
In previous theoretical work\cite{sabass2010efficiency, sabass2012} we have investigated the efficiency of this
surface-driven propulsion since its
exciting potential for use with micro-machines has been amply demonstrated
\cite{ozin2005dream, kline2005catalytic, balasubramanian2011micromachine, baraban2012transport}. 
Chemically driven microswimmers can also be used as model systems to
investigate the physical principles of non-equilibrium diffusive motion 
\cite{golestanian2009anomalous, palacci2010sedimentation, ten2011brownian}. 
Mutual interactions\cite{yang2010swarm, kolmakov2010designing,
kagan2011chemically, Bialke} and the influence of confinement 
on the swimming\cite{qian2008analysis,elgeti2009self, popescu2009confinement}
are a further active field of research.

A number of physical driving mechanisms for hydrogen peroxide
based propulsion have been suggested. One of them is
diffusiophoresis, which is particle motion through its interaction with a
concentration gradient \cite{anderson1989colloid, golestanian2005propulsion, ebbens2011direct}. An alternative, that
is seen to dominate for large-scale swimmers, is the recoil of oxygen bubbles
\cite{gibbs2009autonomously, mei2011rolled}.
Pressure waves were also suggested as a possible origin of motion
\cite{felderhof2010dynamics}.
Finally, it has been found that metallic microswimmers swim through 
self-electrophoresis\cite{paxton2006catalytically, wang2006bipolar}. Self-electrophoresis 
is the swimming of a charged particle through electric interaction with a self-generated,
charged environment. 

The work in this article is motivated by the experiments with metallic swimmers
done by Sen, Mallouk, Wang and co-workers\cite{paxton2006catalytically,
wang2006bipolar} 
following their work with Paxton et al.\cite{paxton2004}. There,
bimetallic microrods (often a platinum-gold particle) were immersed in an
hydrogen peroxide ($H_2O_2$) solution.
The surface of the rods catalyzes a decomposition of $H_2O_2$ into oxygen
$O_2\,(g)$ and water. It has been found that the mechanism for electrokinetic
decomposition of $H_2O_2$ involves an electric current inside the microrod, 
which causes a concentration gradient of ions in the fluid surrounding the
swimmer. Since the rod interacts through an electric potential with the ions, a gradient 
of the latter leads to swimming. The swimming speed increases linearly with the
concentration of $H_2O_2$ until it saturates above concentrations of about 
$5\, \%$. An interpretation of this saturation has
been given in the framework of a Michaelis-Menten-like surface kinetics.
Wang et al. \cite{wang2006bipolar} conducted a comprehensive study using
different combinations of metals for bimetallic microswimmers. 
A clear correlation between the mixed potential difference between the two
metals
and the swimming speed was found. By choice of optimal
materials\cite{laocharoensuk2008carbon} for the anode,
cathode and intermediate part of the swimmer, the speed could even be
increased to more than $50 \,\mu \rm{m/s}$. 
In order to understand the electrokinetic driving mechanism
quantitatively, a few theoretical studies have been conducted. Sundararajan et
al. \cite{sundararajan2008catalytic} compared experimental data with
a numerical model where the rate of the cation exchange was fixed.
Yariv \cite{yariv2010electrokinetic} conducted a thorough study of the case of
a thin diffuse layer of charges surrounding the particle. His linear response
theory 
included surface reaction kinetics explicitly. Moran and Posner
\cite{moran2010locomotion} investigated a full numerical model for 
rather thick, diffuse layers of ions around the swimmer.
Most recently, they included a surface reaction that is second order in $H^+$
concentration\cite{moran2011electrokinetic}. This work provided a closed,
quantitative model that could be directly compared to experimental data.
Their results indicated a quadratic increase of the swimming speed with 
the concentration of $H_2O_2$, which, however, is not observed in the
experiment. 
Building on the work described above, we here investigate a complementary,
non-linear model for
a charged double layer structure that is much thinner than the radius of the
swimmer.

An estimate of the ion concentrations\cite{paxton2006catalytically} around the
swimmers suggests that these
systems are operating quite far away from equilibrium. On the other hand, the
linear speed \textsl{vs.} concentration relation at low and moderate $H_2O_2$
concentrations
does hint that a theory containing linear response elements may still be
valid. Therefore, we choose a combination of numerical modeling and analytical
approximation to capture the non-equilibrium behavior of the system and 
provide simple explanations for qualitative trends.

Any model of self-electrophoretic swimming is hampered by a general lack of
knowledge concerning the complex details of the surface reaction 
mechanism. In particular, the dependence of surface charge accumulation on the
reaction rate is largely unknown. Since the electric potential of the
swimmer directly influences its swimming speed, a better understanding of the
effect of charge screening and ion adsorption is highly desirable. 
To pave the way for further investigations of these problems we model the system
as simple as possible, 
demonstrating only the generic aspects.
\section{The model}
\subsection{Swimmer in an ionic solution}
\begin{figure}[ht]
\begin{center}
\includegraphics[scale=0.73]{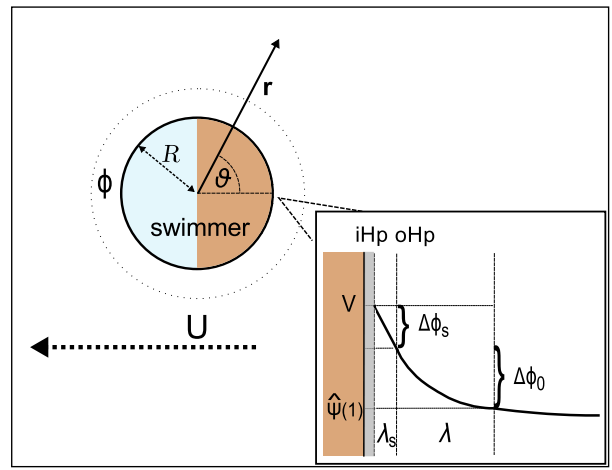}
\caption{\label{fig_1_1}  Schematic representation of the model for a
self-electrophoretic swimmer with
radius $\calR$ moving with swimming speed $U$. The electric potential outside
the swimmer is denoted by $\phi$. The inset shows an enlarged version of our
double layer model. The swimmer surface is covered by an immobile, compact layer
of salt ions, termed Stern layer. The inner Helmholtz plane (iHp) and outer
Helmholtz plane (oHp) delimit the modeled region of the Stern layer with
thickness $\lambda_s$. 
Bordering to the outer Helmholtz plane we have the diffuse layer, where dilute
solutes form a charge cloud around the swimmer. The diffuse layer has a thickness
$\lambda$. }
\end{center} 
\end{figure}

We employ the commonly used hydrodynamic model for electrophoretic effects
\cite{saville1977electrokinetic}. Throughout this publication, we will designate dimensional 
variables with a tilde ( $\tilde{\,}$ ). Dimensional constants are written in
calligraphic letters. The swimmer, a spherical particle with radius $\calR$, is placed in an
infinitely large container (Fig. \ref{fig_1_1}). Swimming speed will be denoted by $\tU$,
The fluid surrounding the particle is assumed to be incompressible and 
Newtonian with a constant viscosity $\iteta$. Mass flow velocity is 
denoted by $\tbv$. The fluid contains ions that carry only one unit charge each. The
concentrations of salt anions and cations in the solution are denoted by
$\tc^{i-}$ and $\tc^{i+}$, respectively.
The hydroxide ($OH^-$) and proton ($H^+$) concentrations, resulting from the
spontaneous dissociation of water and 
from the electrocatalytic process, are denoted by $\tc^{H-}$ and
$\tc^{H+}$, respectively. All types of ions are assumed to
 have the same diffusion constant $\calD$. Concentrations of cations and anions
are denoted summarily by 
\begin{align}
\begin{split}
 \tc^{+} & \equiv\tc^{H+}+\tc^{i+},\\
\tc^{-} & \equiv\tc^{H-}+\tc^{i-}.
\end{split}
\end{align}
The ions couple to an electric field $\tilde{\phi}$. In the bulk, far away from the swimmer, 
we have charge neutrality and the concentrations of ions are 
equal $\tc^-(\infty)=\tc^+(\infty)$.
The swimmer is axially symmetric. Therefore we use a
spherical coordinate system aligned in the $\mathbf{\hat{e}}_z$ direction, where
$\tr$ is the 
distance from the particle center and $\vartheta$ is the inclination angle. The
unit vectors of the spherical
system are denoted by $ \mathbf{\hat{e}}_r, \,  \mathbf{\hat{e}}_{\vartheta}$. 

We non-dimensionalize length with the swimmer radius $\mathcal{R}$ and the
concentrations of 
ions $\tc^{\pm}$ with the bulk concentration of salt ions
$\mathcal{C}=\tc^{i,+}(\infty)=\tc^{i,-}(\infty)$. Energies are normalized with
the
thermal energy scale $\calkT$. Accordingly, the electric potential $\tphi$ is made
non-dimensional with the thermal voltage $\calkT/Z \ite \simeq 25\,\rm mV$. Throughout the publication we set $Z=1$.
Diffusive fluxes are non-dimensionalized with $\calD \calC/\calR$.
The velocity scale of the flow $\tbv$ and the swimming speed $\tU$ is given by a typical speed $\calU$.
The Pecl\'et number associated with transport of cations and anions is given by
$\Pe\equiv \calU \calR/\calD$.
\subsection{Poisson-Boltzmann and diffusion equations}
The fluxes of cations and anions are given in dimensionless form by
\begin{align}
 \bj^{\{H,i\},+} &=-  \nabla c^{\{H,i\},+} - c^{\{H,i\},+}\nabla \phi +\Pe \bv
c^{\{H,i\},+},\label{eq_jp}\\
 \bj^{\{H,i\},-} &=-  \nabla c^{\{H,i\},-} + c^{\{H,i\},-}\nabla \phi +\Pe \bv
c^{\{H,i\},-}\label{eq_jm}.
\end{align}
Diffusion equations guarantee the conservation of substance in the fluid. They
determine the concentration field and read
\begin{align}
\nabla\cdot \bj^{\{H,i\},+} &=0 \label{eq_diff_cp},\\
\nabla\cdot \bj^{\{H,i\},-} &=0\label{eq_diff_cm}.
\end{align}
The Poisson-Boltzmann equation provides a connection between the ion density
and the electric field. It reads in dimensionless form
\begin{align}
 \nabla^2 \phi &= -\frac{1}{2 \,\lambda^2}(c^+ - c^-).\label{eq_PB}
\end{align}
The lengthscale of the potential $\phi$ in Eq. (\ref{eq_PB}) is given by the
dimensionless Debye length $\lambda$, which we define as
\begin{equation}
 \lambda \equiv \sqrt{\frac{\iteps \,\calkT}{8 \pi\, \ite^2 \,\calNa\calC}}\,
\frac{1}{\calR} \label{eq_debye_len}
\end{equation}
with $\calNa$ being the Avogadro constant and $\iteps$ the permittivity of the
solution.
\subsection{Stern layer and boundary conditions}
In the bulk, far away from the swimmer, the electric potential takes on an
uniform value which we set
to zero
\begin{align}
\phi(r \rightarrow \infty,\vartheta) &= 0.
\end{align}
The presence of salt in the solution suggests some form 
of adsorption of ions onto the metal surface. The inset of Fig. \ref{fig_1_1} illustrates the employed 
model of a Stern (compact) -layer of immobile ions covering the metal surface 
\cite{delgado2005measurement}. All the charge in the Stern layer is assumed to be concentrated in one plane, called inner Helmholtz plane (iHp). 
This plane almost coincides with the metal surface. We do not consider the details of the
structure beyond the inner Helmholtz plane. The boundary between the Stern
layer and the diffuse layer is termed
outer Helmholtz plane (oHp). 
The voltage difference between inner and outer Helmholtz plane is called Stern
layer voltage 
drop $\Delta \phi_s(\vartheta)$. The electric potential at the inner Helmholtz
plane $V$ equals
 $\Delta \phi_s(\vartheta)$ plus the change of the potential between outer
Helmholtz plane and infinity $\phi(1,\vartheta)$:
\begin{align}
V = \Delta \phi_s(\vartheta) + \phi(1,\vartheta).\label{eq_phi_stern_condition}
\end{align}
Since we have an electric current flowing through the swimmer, the potential
$V$ may not be strictly constant. However, the electric conductivity
of the metal swimmer is much higher than the conductivity of the surrounding
solution.
We therefore assume that the variation of $V$ is negligible. \textsl{A priori}, the potential
change over the Stern layer, $\Delta\phi_s(\vartheta)$, is unknown. If it depends on the
potential right outside the immobile layer, it becomes $\vartheta$-dependent. Since one
postulates that the Stern layer outside the inner Helmholtz plane is charge free, Gauss's law
can be employed to arrive at 
\begin{equation}
\Delta\phi_s(\vartheta) = -\lambda_s \frac{\p \phi(r,\vartheta)}{\p r}|_{r =1},
\label{eq_determine_phi_s}
\end{equation}
where the parameter $\lambda_s$ is the dimensionless thickness of the Stern
layer between inner and outer Helmholtz plane. 

The electrocatalytic process leading to a decomposition of $H_2O_2$ is 
supposed to take place at the inner Helmholtz plane. It leads to an absorption
and emission of $H^+$ ions with 
a rate $\alpha(\vartheta)$, non-dimensionalized with the flux
$\calC\,\calD/\calR$. Boundary conditions, determining the  
concentrations of ions, result as
\begin{align}
 \hbe_r\bj^{H,+}(1,\vartheta) &= \alpha(\vartheta),\\
 \hbe_r\bj^{i,\pm}(1,\vartheta) &=  \hbe_r\bj^{H,-}(1,\vartheta)=0,\\
c^{i,\pm}(r \rightarrow \infty,\vartheta) &= 1,\\
c^{H,\pm}(r \rightarrow \infty,\vartheta) &= \delta.
\end{align}
The constant $\delta$ is the relative concentration of $H^+$ ions in the bulk. It is defined by
\begin{equation}
 \delta \equiv \tc^{H,+}(\infty)/\calC.
\end{equation}
In order to keep the charge of the swimmer constant the exchange of cations at
its surface must 
satisfy
\begin{equation}
 \int \hbe_r \bj^{H,+}(1,\vartheta)\, \rmd A_{r=1} = 0.
\label{eq_charge_conservation}
\end{equation}
\subsection{Hydrodynamic equations}
The mass flow velocity $\bv$ is determined by the Stokes equation for overdamped
motion 
\begin{align}
\nabla\cdot\bs=\nabla^2 \bv - \nabla p = g\left(c^+ - c^-\right)\nabla\phi
\end{align}
where $p$ is the pressure and $\sigma \equiv \nabla \bv+(\nabla \bv)^{T} -p
\mathbf{I}$ is the hydrodynamic stress tensor. 
We have also defined the non-dimensional constant 
\begin{equation}
 g \equiv \frac{\calR \, \calC \calkT}{\iteta\, \calU}.
\end{equation}
The boundary conditions on the fluid flow are 
\begin{align}
\bv(r \rightarrow \infty,\vartheta) = \bU,\\
\bv(0,\vartheta) = 0,
\end{align}
where the swimming velocity $\bU$ is calculated from the balance of forces on
the
particle
\begin{align}
0= \int g\left(c^+ - c^-\right)\nabla\phi\, \rmd V +\int \bs \hbe_r \rmd
A_{r=1}.\label{eq_force_balance}
\end{align}
For a body force $g\left(c^+ - c^-\right)\nabla\phi$ that is
independent of the fluid velocity $\bv$ one can employ
Teubner's formula\cite{teubner1982motion} to calculate the swimming speed as
\begin{equation}
\begin{split}
  U   =  -\frac{g}{6 \pi}\int 
[(\frac{3 }{2 r} - \frac{1}{2 r^3}
-1)\cos\vartheta\,\left(c^+ - c^-\right)\p_r\phi \\
-(\frac{3 }{4 r} + \frac{1}{4 r^3}
-1)\sin\vartheta\,\left(c^+ - c^-\right)\frac{\p_{\vartheta}\phi}{r}]\,
\mathrm{d}V
\label{eq_teubner_U}.
\end{split}
\end{equation}

\subsection{Redox reactions at the surface of the swimmer}\label{sec_rates}
\begin{figure}[ht]
\begin{center}
\includegraphics[scale=0.73]{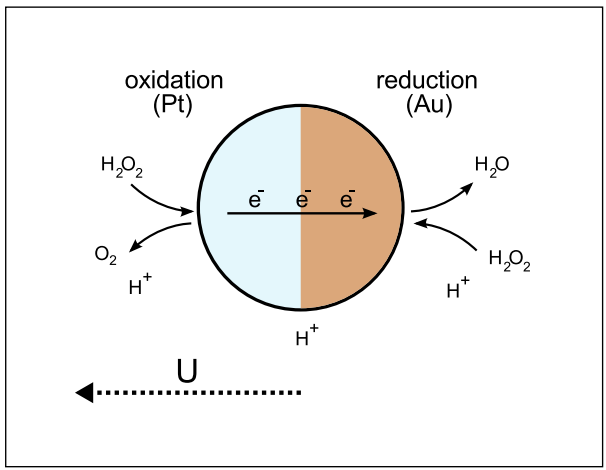}
\caption{\label{fig_1_2}Sketch of the electrokinetic process taking place at the
swimmer's surface as described in Sec. \ref{sec_rates}. 
Note that the swimmer is found to move with the oxidizing end forwards.}
\end{center} 
\end{figure}

The electrochemical reactions leading to the decomposition of
$H_2O_2$ at the swimmer's surface are complicated, especially since a number of
inhibitory processes are involved. Experimental evidence supports the view
 that the reaction rate depends non-linearly on the $H_2O_2$ concentrations
\cite{paxton2005motility, howse2007self}, most visibly for $c_{H_2O_2} \gtrsim
5\,\%$. The inferred (Michaelis-Menten-like) kinetics for a microswimmer agrees
with other
findings\cite{hall1997electrochemical}. This kinetics is probably due to
saturation of
active sites at the surface. Here we
do not model the saturation but focus on low concentrations of educts and
products. Experiments with comparatively large (mm-sized) electrodes suggest
that the decomposition 
of hydrogen-peroxide at metal surfaces in acidic medium happens via a formation 
of oxides at the surfaces\cite{bianchi1962catalytic}. As also discussed in
App.
\ref{sec_app_reaction}, this intermediate may serve to justify an effective
description in
terms of a first order surface reaction. Fig. \ref{fig_1_2} illustrates that
protons combine with $H_2O_2$ for the 
reduction of $H_2O_2$. Therefore, the rate of $H^+$ consumption is assumed to 
be $\sim c^{H,+}(1,\vartheta) c_{H_2O_2}$. Electrocatalytic oxidation is
supposed to supply the protons with a rate proportional to $c_{H_2O_2}$. 

The redox reactions at the swimmer's surface cause proton fluxes and involve a
charge
transfer. Therefore, the electric potential influences the reaction rate. We
model
this dependence with the established Butler-Volmer
kinetics\cite{newman2004electrochemical,bonnefont2001analysis,
biesheuvel2009imposed}. 
It is assumed that the reaction takes place at the inner Helmholtz plane, almost
coinciding 
with the metal surface of the swimmer (see Fig. \ref{fig_1_1}).
Since the Stern layer is only described in an effective way, the energetic 
cost of passing from the outer to the inner Helmholtz plane 
must be incorporated explicitly into the reaction rates. This is commonly done
by introducing rates that depend
exponentially on $\Delta\phi_s(\vartheta)$ and satisfy detailed balance. We
choose rates $\sim e^{\pm\frac{\Delta
\phi_s(\vartheta)}{2}}$, where the transfer coefficient, modifying $\Delta
\phi_s$ in the exponents, is set to $1/2$.
For simplicity, we neglect further direct effects of the compact salt layer,
that could, e.g., reduce the
catalytic activity by covering the active sites. Evidently, $\phi$ also modifies
the reaction rate by influencing the 
concentration of protons right outside the Stern layer $c^{H,+}(1,\vartheta)$.
This dependence is however implicit and it does not appear
in the rate equation. The overall rate of proton
exchange at the swimmer's surface takes the form
\begin{align}
\begin{split}
 \alpha(\vartheta) = -\left(K_1 +k_1 \cos\vartheta\right) e^{-\frac{\Delta
\phi_s(\vartheta)}{2}} c^{H,+}(1,\vartheta) c_{H_2O_2}  \\
+\left(K_2- k_2\cos\vartheta\right)e^{\frac{\Delta
\phi_s(\vartheta)}{2}}c_{H_2O_2} \label{eq_alpha}, 
\end{split}
\end{align}
with $K_1 \geq k_1$ and $K_2 \geq k_2$. The first term in Eq. (\ref{eq_alpha})
models the catalytic reduction of 
$H_2O_2$ while the second term models the provision of $H^+$ ions through an
oxidation. 
The constants $k_1$ and $k_2$ reflect the difference of redox potentials between
the two metals composing the swimmer.
Note that the linear dependence of $ \alpha(\vartheta)$ on the $H_2O_2$
concentration in the bulk merely
rescales all rate constants in the same way. The bulk concentration of $H_2O_2$
is a
model parameter. We choose to express $c_{H_2O_2}$ in (dimensionless) 
fractions of $1\,\%$ wt/v in order to facilitate the comparison with
experimental data. 
In not modeling the $H_2O_2$ concentration explicitly we have assumed that a 
diffusive supply of this reagent is not the bottleneck of the reaction mechanism
and that
the encounter of $H_2O_2$ with the surface can be incorporated in the effective
rate constants.

\section{Theory for thin diffuse layers}
If the Debye length is small, one can attempt an expansion of the concentration
and potential fields in orders of $\lambda$, which largely simplifies the nonlinear
equations. We follow here the classical approach outlined, e.g, by Hunter
\cite{hunter1987foundations} which was used recently in the group of 
Bazant \cite{bonnefont2001analysis, chu2006nonlinear} to model time dependent 
electrochemical processes. 
\subsection{The nonlinear problem}
The technical details of the nonlinear double layer model are lengthy. We
therefore
defer them to App. \ref{sec_app_thin_layer}. Here we only sketch the
mathematical procedure and present the result for the swimming speed. 
The field variables are expanded in powers of $\lambda$ as
\begin{align}
\begin{split}
c^{i,\pm} &=c^{i,\pm}_0 +\lambda \,c^{i,\pm}_1 +\ldots,\\
c^{H,\pm} &=c^{H,\pm}_0 +\lambda \,c^{H,\pm}_1 +\ldots,\\
\phi &= \phi_0 +\lambda \, \phi_1 +\ldots\,.
\label{eq_def_phi_expand}\\
\end{split}
\end{align}
In the following, we only consider the terms of lowest order, $O(\lambda^0)$.
The strategy is now to solve the non-linear field equations both in an
outer region, far away from the surface, and inside the diffuse layer. 
The inner and outer solutions are then matched to each other, such that an
uniformly valid solution can be constructed. 
We distinguish the outer variables from the inner ones through a hat
($\,\hat{\,\,}$) on the former. Also, the potential $\phi_0$ is written as sum of an inner and outer solution
\begin{equation}
\phi_0 = \psi_0 + \hpsi_0.
\end{equation}
The quasi one-dimensional equations for the inner variables can be solved analytically while the
 outer equations must be solved numerically. The result for the electric
potential in the diffuse layer is 
\begin{equation}
\begin{split}
\phi_0(y,\vartheta) =&  \psi_0(y,\vartheta) + \hpsi_0(1,\vartheta)= \\
&2 \ln\left[\frac{1+\gamma(\vartheta) e^{-\sqrt{B(\vartheta)}
y}}{1-\gamma(\vartheta) e^{-\sqrt{B(\vartheta)} y}}\right] +
\hpsi_0(1,\vartheta), \label{eq_inner_psi0} 
\end{split}
\end{equation}
where we defined $y \equiv (r-1)/\lambda$. The function $\gamma(\vartheta)$ is
determined by the boundary conditions on 
the swimmer surface (see App. \ref{sec_app_thin_layer}). $B(\vartheta)$ is
proportional to the overall concentration of ions at the outer boundary of the
diffuse layer 
\begin{equation}
B(\vartheta)\equiv
{\left(\hc_0^{H, +}(1,\vartheta)+\hc_0^{H, -}(1,\vartheta)+\hc_0^{i,
+}(1,\vartheta)+\hc_0^{i, -}(1,\vartheta)\right)}/2.
\end{equation}
Given (numerical) solutions for $\hpsi_0$ and the potential at the swimmer
surface $V_0$,
we can use Eq. (\ref{eq_phi_stern_condition}) to calculate the potential drop
across the diffuse layer $\Delta \phi_0(\vartheta)$ as
\begin{equation}
  \Delta \phi_0(\vartheta)  = V_0 - \Delta
\phi_s(\vartheta) - \hpsi_0(1,\vartheta), \label{eq_stern_bc_main}
\end{equation}
where the voltage drop across the Stern layer $\Delta
\phi_s(\vartheta)$ is calculated through Eq. (\ref{eq_determine_phi_s}). $\Delta
\phi_0(\vartheta)$ can be interpreted as 
an angle-dependent zeta potential. The final result for the swimming speed reads, 
in terms of the above defined variables,
\begin{equation}
\begin{split}
 U \approx \,& U_0 =  \, g\lambda^2\int_0^{\pi}\,[ \Delta \phi_0(\vartheta)\,
\frac{\p
\hpsi_0(1,\vartheta)}{\p \vartheta}\,\sin^2(\vartheta) - \\
&4 \ln\left(\cosh\left(\frac{\Delta\phi_0(\vartheta)}{4}\right)\right)  
\frac{\p \ln\left( B(\vartheta) \right)}{\p \vartheta}\,\sin^2(\vartheta)\,]\rmd
\vartheta \label{eq_U}.
\end{split}
\end{equation}
We will omit the index at $U_0$ in the following since we do not deal with
higher orders
in $\lambda$. Remarkably, Eq. (\ref{eq_U}) has the same form as the celebrated
formula by 
Dukhin and Derjaguin \cite{dukhin1974surface, anderson1989colloid}, which
was derived for the linear response regime. The only difference is that the 
potential drop across the diffuse layer can not be approximated by the
equilibrium zeta potential and is now $\vartheta$-dependent. 
The dependence of $U$ on $\Delta \phi_0$ allows to separate
electrophoretic and diffusiophoretic contributions in the lowest order swimming
speed\cite{prieveElectrolytes}. The first
term in the integrand of Eq. (\ref{eq_U}) changes sign with $\Delta \phi_0$, similar to
electrophoresis, where the particle swimming is also reversed with the sign of the
zeta potential. Therefore, the first term can be interpreted as the electrophoretic part. 
The second term does not change sign with the potential $\Delta
\phi_0$ since $-4\ln\left(\cosh\left(\Delta \phi_0/4\right)\right)=2
\ln(1-\gamma^2)$ is negative semi-definite. Its dependence on a 
gradient in overall concentration suggests an interpretation of the second term as
diffusiophoretic contribution.
\subsection{Analytical approximation} \label{sec_approx_U}
We assume now that the redox potential difference between the two sides of the
swimmer is considerably smaller than its overall redox potential. This case can
be modeled by setting $K_1,\, K_2 \gg k_1,\,k_2$. In a steady state, the
constant $H_2O_2$ decomposition will modify the electric charge
of the swimmer. If, additionally, $k_1$ and $k_2$ are neglected, no net 
cation flux can take place since a radially symmetric flux violates the charge 
conservation. The concentration of protons near the surface then follows a 
Boltzmann distribution $\sim\delta \exp\left[-\phi_{0,0}(y)\right]$, where the second
index $0$, here at $\phi_{0,0}$, indicates that we are considering the solution
of order $(k_{1}^0,\,k_{2}^0)$.
In this limit, Eq. (\ref{eq_alpha}) becomes
\begin{align}
 0 = -K_1 e^{-\frac{\Delta \phi_{s,0}}{2}} \delta\, e^{-\phi_{0,0}(0)}
c_{H_2O_2}  +K_2e^{\frac{\Delta \phi_{s,0}}{2}}c_{H_2O_2},
\label{eq_lowest_order_alpha}
\end{align}
It follows that the swimmer's potential in steady state
\begin{align}
V_{0,0} \equiv \Delta \phi_{s,0} + \phi_{0,0}(0) = \ln\left(K_1
\delta/K_2\right)
\label{eq_lowest_order_V}
\end{align} 
does not depend explicitly on the $H_2O_2$ concentration to lowest order. 
The swimming speed is calculated in App. \ref{sec_app_approx}. We find
\begin{align}
U \approx g \lambda^2 \, \frac{2 \ln\left(1+\gamma_0\right)}{1+\delta} \int
\alpha_{0,1} \cos\vartheta \,\rmd\cos\vartheta \label{eq_lin_resp_U}
\end{align}
where the integral is a measure of the dipole moment of the first order $H^+$
emission/absorption rate $\alpha_{0,1}$. We also defined $\gamma_0 \equiv
\tanh(\phi_{0,0}(0)/4)$. Eq. (\ref{eq_lin_resp_U}) is similar to previously
derived linear response formulas\cite{yariv2010electrokinetic}. When inserting
$\alpha_{0,1}$, which is calculated in App. \ref{sec_app_approx}, we have
\begin{equation}
U \approx g \lambda^2 \frac{8}{3} \frac{\ln\left(1+\gamma_0\right)}{1+\delta}\, 
\frac{\delta \left(k_2 + k_1 \frac{K_2}{K_1}\right)\,c_{H_2O_2} }{2 \delta \,
e^{-\frac{\Delta \phi_{s,0}}{2}} + K_2 c_{H_2O_2}}. \label{eq_approx_U}
\end{equation}
For a weak potential outside the compact layer, $\phi_{0,0} \lesssim 1$, one can
estimate the dependence of the swimming speed on the particle potential through
$\ln\left(1+\gamma_0\right) \simeq (V_0-\Delta \phi_{s,0})/4$.

\section{Numerical values for the model
constants}\label{sec_mall_c_pert_appendix}
The thermal energy scale is $\calkT = 4.11\times 10^{-21}
\rm{J}$. For the solution viscosity, we choose $\eta = 9\times 10^{-4}\,
\rm{Pa\,s}$. For the permittivity of the solution around the swimmer, we
choose
$\iteps = 80\, \iteps_{vak}$. Typically, the experimental swimmers have lengths in the 
$300\, \rm{nm}$ - $5\, \mu\rm{m}$ range. We set the radius of our 
swimmer to
\begin{equation}
 \calR \equiv 1\, \mu\mathrm{m}.
\end{equation}
A velocity scale can be defined, which is independent of the concentration scale. We set
\begin{equation}
\calU g \lambda^2 = \frac{(\calkT)^2\, \iteps}{8 \pi \iteta \, \ite^2
\calR}\simeq 20.6\,\mu\rm{m}/\rm{s}.
\end{equation}
In this work we rely on the addition of extra salt to the solution to explicitly
justify the usage
of the thin diffuse layer model. Where not explicitely mentioned otherwise, we
set the bulk concentration of salt ions in the following to
\begin{equation}
\calC = 5\times10^{-5}\;\mathrm{mol/L}.
\end{equation}
For a monovalent salt, e.g., $NaNO_3$ as used by
Paxton et al. \cite{paxton2006catalytically}, according to Eq.
(\ref{eq_debye_len}), we have a thickness of the diffuse layer of $\simeq 12\, \rm{nm}$.
Since this is much smaller than $\calR = 1\,
\mu\rm{m}$ our theory for $\lambda \ll 1$ is expected to work well.
The absolute thickness of the Stern layer
is usually assumed to be in the order of a molecular
diameter\cite{wang2011accurate}, which might be 
about $0.3 \,{nm}$.

For propulsion of our microswimmers we consider a $\left[10^{-4}\ldots 5\right]\%$  
solution of $H_2O_2$ in water. Percent of $H_2O_2$ concentration are given in wt/v (1 \% wt/v corresponds to 10g $H_2O_2$
per liter solvent, molar mass of $H_2O_2$: 34.02 g/mol). The pH value decreases
here almost linearly with an increase of $H_2O_2$
concentration\cite{USperoxide}. It lies roughly between 7 and 5.
This sets the
concentration of bulk protons and hydroxide ions to $\tc^{H, \pm}(\infty)
\simeq \left[10^{-7}\ldots10^{-5}\right]\;\mathrm{mol/L}$. The relative
concentration of $H^+$ ions in 
the bulk is then in the range $\delta = \tc^{H}(\infty)/\calC \simeq
\left[10^{-3}\ldots10^{-1}\right]$. 

For the diffusion constant, we choose $\calD = 7 \times 10^{-9} \rm{m/s}$.
The ion fluxes at the surface are then non-dimensionalized with 
\begin{equation}
\calC \calD/\calR = 3.5\times 10^{-4}\,\rm{mol/m^2\,s}.
\end{equation}
A measured\cite{paxton2006catalytically} evolution rate of $O_2$ for a $3.7\,\%$ solution of $H_2O_2$ is
 $8.7 \times 10^{-6} \rm{mol/m^2\,s}$. Within our reaction model, Eq. (\ref{eq_alpha}), this rate
roughly determines the magnitude of the second term, responsible for the oxidation of $H_2O_2$. 
For $c_{H_2O_2} = 1$ ($1\,\%$) we estimate that $K_2\, c_{H_2O_2}$ is on the
order of $10^{-2}$. We assume that its value scales linearly in the
range of employed $H_2O_2$ solutions. The constants $K_1,\,k_1,\,k_2$ are chosen
such that the we obtain reasonable values for the swimming speed and the proton
fluxes.
\section{Results}
In order to facilitate a comparison with experimental data, we often display
dimensional quantities in the following. The dimensional speed is calculated as
$\tU =
\calU\,U = 20.6 \,\mu\rm{m/s} \times U /(g \lambda^2)$. Results for the analytical
approximation of $U$ (Eq. (\ref{eq_approx_U})) are also shown where applicable.  
\subsection{Reaction induced concentration distortion}
\begin{figure}[ht]
\begin{center}
\includegraphics[scale=0.67]{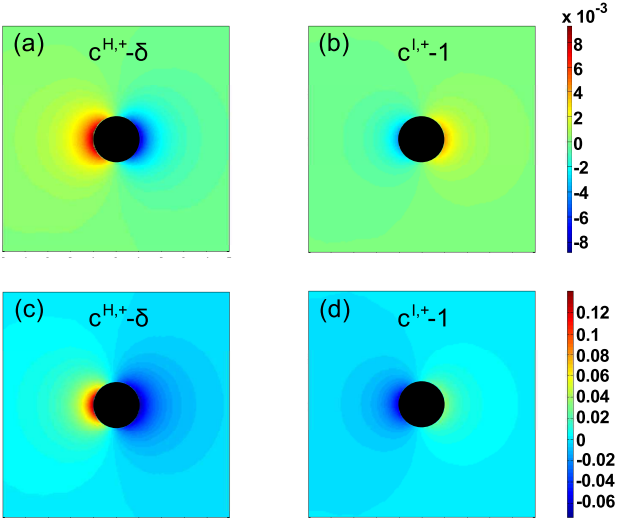} 
\caption{\label{fig_2}
Colorcoded concentrations of $H^+$ ions and positive salt ions. Figs. a,b
and c,d each have have the same color coding. $c_{H_2O_2} =1$ ($1\,\%$), $K_1 =
0.0185$, $K_2 = 0.185$, $\delta = 0.1$, and $\lambda_s = 0$. For Figs. a,b:
$k_1 =0.1\,K_1 $, $k_2=0.1\,K_2$
resulting in $V_0=-4.61$ and
$U=-0.0398$. For Figs. c,d: $k_1=\,K_1$, $k_2=\,K_2$ resulting in
$V_0=-5.03$ and $U=-0.485$.
}
\end{center} 
\end{figure}

Fig. \ref{fig_2} shows exemplary concentration fields of positive ions around
the swimmer. While the concentrations of $H^+$ in Figs. \ref{fig_2} a,c are directly
determined
through the surface reactions, the inhomogeneous distribution of positive salt
ions seen in Figs. \ref{fig_2} b,d is an indirect effect.
The positive salt ions accumulate oppositely to the $H^+$ ions.
They present a non-negligible contribution to the local charge balance.
In Figs. \ref{fig_2} a,b we have $(k_1,\, k_2) \ll (K_1,\, \,K_2)$, i.e, the
decomposition of $H_2O_2$ happens fairly homogeneously around the swimmer. In this
case the concentration profile is antisymmetric and follows the distribution of
surface reactivity. In Figs. \ref{fig_2} c,d we have set $k_1 = K_1$ and
$k_2=K_2$, i.e., the reaction happens mainly in an asymmetric
way, where oxidation and reduction take place on different sides of the swimmer.
Also, the concentration of $H^+$ ions in the bulk is not assumed to be
excessive. Then the concentration 
variation around the swimmer feeds back on the overall reaction
rate, Eq. (\ref{eq_alpha}). Consequently, the antisymmetric distribution of the
cations 
seen in Figs. \ref{fig_2} a,b is lost. This reaction induced concentration
distortion is a direct consequence of the usage of the mass action law to model
the surface reaction. In
our previous work we have found that it is important for the calculation of the
swimming speed\cite{sabass2012}.
\subsection{Swimming and $H_2O_2$ concentration, fixed pH}
\begin{figure}[ht]
\begin{center}
\includegraphics[scale=0.63]{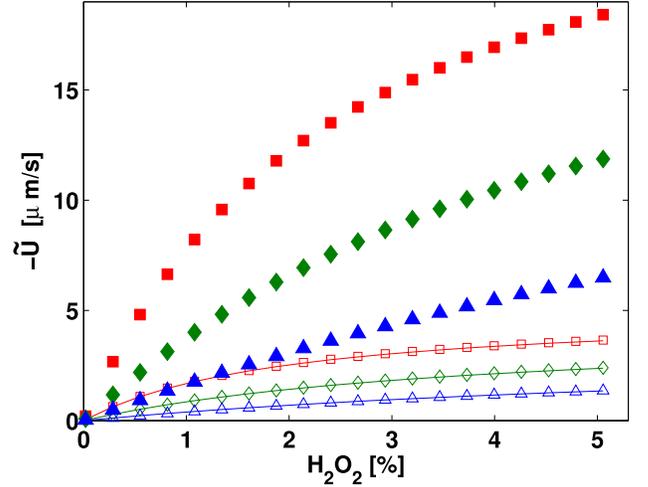}
\caption{\label{fig_3_1}
Swimming speed $\tU$ \textsl{vs.} $H_2O_2$ concentration at fixed pH value.
$K_1= 0.005$,
$\delta= 0.1$, $\lambda_s = 0$. Symbols are numerical results. Full lines are
approximative speeds from Eq. (\ref{eq_approx_U}). (Red box): $K_2 = 0.1$, $k_2
= K_2$, $k_1 = K_1$. (Open red box): $K_2 = 0.1$, $k_2 = 0.25\,K_2$, $k_1 =
0.25\,K_1$. (Green diamond): $K_2 = 0.05$, $k_2 = K_2$, $k_1 = K_1$. (Open green
diamond): $K_2 = 0.05$, $k_2 = 0.25\,K_2$, $k_1 = 0.25\,K_1$.(Blue triangle):
$K_2 = 0.025$, $k_2 = K_2$, $k_1 = K_1$. (Open blue triangle): $K_2 = 0.025$,
$k_2 = 0.25\,K_2$, $k_1 = 0.25\,K_1$.
}
\end{center} 
\end{figure}

Fig. \ref{fig_3_1} demonstrates the dependence of the swimming speed $\tU$ on the
concentration
of $H_2O_2$ with a fixed pH value. A fixed pH value means that the availability
of $H^+$ in the bulk, determined by $\delta$, does not change with $H_2O_2$
concentration. Eq. (\ref{eq_approx_U}) approximates the numerically calculated
swimming speed well for $k_1 < K_1$ and $k_2 < K_2$. However, the condition $k_2
<K_1$ for the validity of the analytical approximation seems here unnecessary. 

\begin{figure}[ht]
\begin{center}
\includegraphics[scale=0.63]{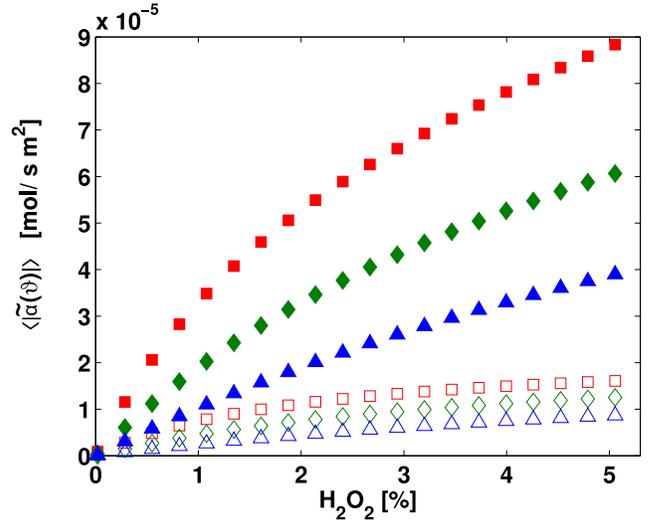}
\caption{\label{fig_3_2}
Average ion current $\langle |\talpha(\vartheta)| \rangle$ \textsl{vs.}
$H_2O_2$ concentration at fixed pH value. Symbols are the same as in Fig.
\ref{fig_3_1}.
}
\end{center} 
\end{figure}

\begin{figure}[ht]
\begin{center}
\includegraphics[scale=0.63]{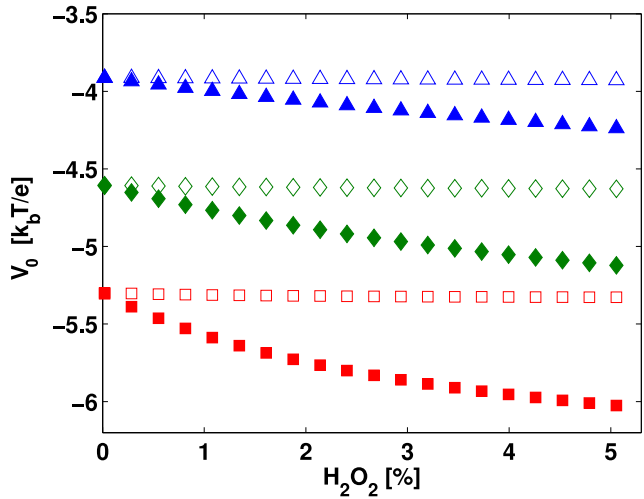}
\caption{\label{fig_3_3}
Electric potential of the swimmer $V_0$ \textsl{vs.} $H_2O_2$
concentration at fixed pH value. Symbols are the same as in Fig. \ref{fig_3_1}.
}
\end{center} 
\end{figure}

The curves in Fig.
(\ref{fig_3_1}) have a concave shape, which is not attributed to a saturation of
the
catalytic surface in our model. Rather, it results
from a limitation of the reaction rate through the diffusion around the 
swimmer. Quantitative understanding of this effect can be gained from the
approximative Eq. (\ref{eq_approx_U}). Here we find a leveling off of the
swimming speed since the rate of ion flux through the swimmer saturates for $
K_2 c_{H_2O_2}
\gtrsim 2 \delta \,e^{-\frac{\Delta \phi_{s,0}}{2}}$. This estimate is seen to
hold also for the results from the
nonlinear numerics. A negative Stern layer voltage drop $\Delta \phi_s$ shifts
the saturation of the speed to higher values of $c_{H_2O_2}$, while also reducing the speed (see Sec.
\ref{sec_stern_effect}).
Figs. \ref{fig_3_2}, \ref{fig_3_3} show how the average reaction rate
$\langle|\alpha(\vartheta)|\rangle$ and the electric potential on the swimmer's
surface $V_0$ change with the increase of $H_2O_2$. Note that
$\langle|\alpha(\vartheta)|\rangle$ can not be
identified with the measurable oxygen evolution rate.
$\langle|\alpha(\vartheta)|\rangle$ 
vanishes for equal redox potentials on both sides of the swimmer, i.e., if
$k_1=k_2=0$, while the oxygen evolution rate in this limit is, within our model,
an
unknown constant. 

\subsection{Swimming and $H_2O_2$ concentration, variable pH}
\begin{figure}[ht]
\begin{center}
\includegraphics[scale=0.63]{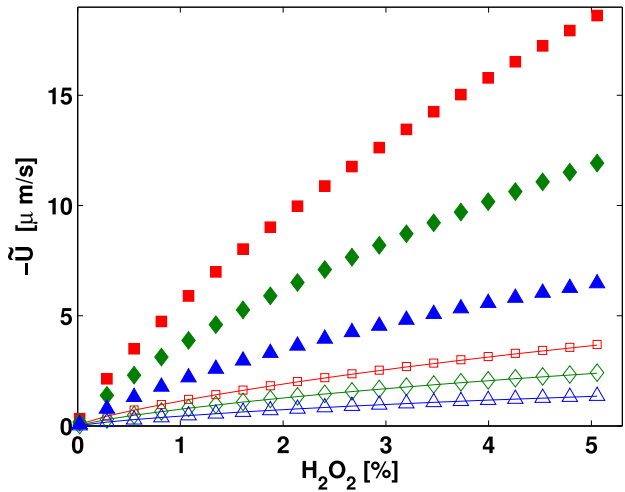}
\caption{\label{fig_3_4}
Dependence of swimming speed $\tU$ on $H_2O_2$ concentration where the pH value
varies
with $c_{H_2O_2}$. We use $\delta = 0.002 + 0.02\,c_{H_2O_2}$. Symbols are the
same as in Fig. \ref{fig_3_1}.
}
\end{center} 
\end{figure}
Since the concentration of $H_2O_2$ also influences the pH value we plot in Fig.
\ref{fig_3_4} the speed \textsl{vs.} $c_{H_2O_2}$ with $H_2O_2$-dependent $\delta$. 
The pH value decreases almost linearly with an increase of $H_2O_2$
concentration\cite{USperoxide}.
Therefore we set $\delta = \delta_0 + \delta_1 \,c_{H_2O_2}$ where the
dissociated water molecules at a pH value of 7 contribute $\delta_0 =
(10^{-7} \rm{mol/L})/\calC$. The slope of the function is roughly estimated as $\delta_1
=0.02$. The reaction rate is not expected to saturate here as seen from Eqns. (\ref{eq_approx_U},\ref{eq_a_first_order}). 
The speed becomes an almost linear function of the $H_2O_2$ concentration. A remaining concave tendency of the functions is
due to a reduction of the Debye length and a change of the particle potential with the pH value. 
The curves in Fig. \ref{fig_3_4} are very similar to what is measured experimentally \cite{paxton2005motility, laocharoensuk2008carbon}. 

\subsection{Swimming and $H_2O_2$ reduction rate}
\begin{figure}[ht]
\begin{center}
\includegraphics[scale=0.63]{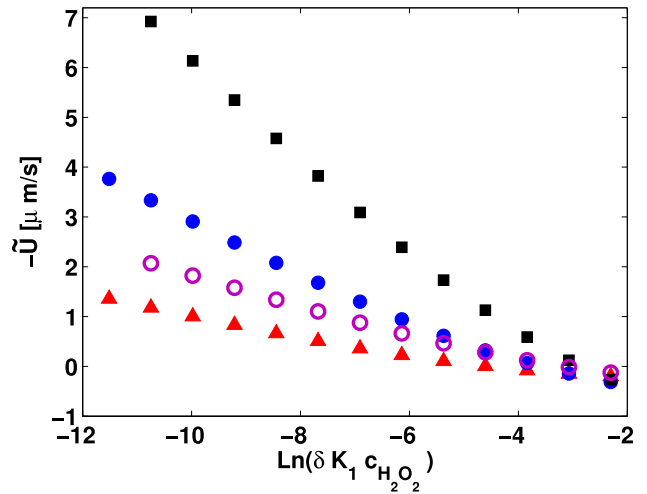}
\caption{\label{fig_4}
Dependence of swimming speed $\tU$ on the rate constant of $H_2O_2$ reduction
$K_1$
at $c_{H_2O_2}=1$ ($1\,\%$) with $k_2 = K_2$, $k_1 = K_1$, and $\lambda_s = 0$.
(Black
box): $\delta= 0.1$, $K_2 = 0.05$. (Blue Dot): $\delta= 0.1$, $K_2 = 0.025$.
(Violet circle): $\delta =0.01$, $K_2 = 0.025$. (Red triangle): $\delta= 0.1$,
$K_2 = 0.01$.
}
\end{center} 
\end{figure}
In Fig. \ref{fig_4} we investigate the role of the rate constants of $H_2O_2$
reduction, $K_1$ and $k_1$,
for the swimming speed. We keep the quotient of $K_1$ and $k_1$ constant and
Eq. (\ref{eq_approx_U}) therefore suggests
that the swimming speed is only influenced through its dependence on the
particle potential $V_0$ via the 
function $\gamma$. The approximation for the swimmer's potential, Eq.
(\ref{eq_lowest_order_V}), in turn suggests that $V_0$ 
depends logarithmically on $K_1\delta/K_2$. The numerical data qualitatively
supports
a scaling of $\tU$ with this logarithm. The magnitude of 
the swimming speed is seen to decrease with an increase of $K_1$. The negative sign of $\tU$ is in 
accordance with the experimental finding that the swimming occurs in direction of the
$H_2O_2$-oxidizing end \cite{wang2006bipolar}. However, for $K_1\delta > K_2$ we also find a 
reversion of the swimming direction, which results from the sign change of $V_0$. 

\subsection{Effect of the Stern layer}\label{sec_stern_effect}
\begin{figure}[ht]
\begin{center}
\includegraphics[scale=0.63]{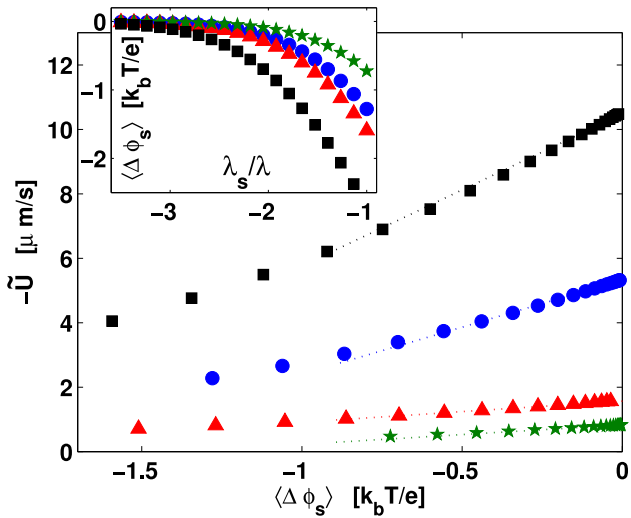}
\caption{\label{fig_5}
Dependence of swimming speed $\tU$ on the average Stern layer voltage drop
$\langle\Delta \phi_s(\vartheta)\rangle$. Inset: dependence of $\langle\Delta
\phi_s(\vartheta)\rangle$ on relative Stern layer thickness
$\lambda_s/\lambda$. 
$c_{H_2O_2}=1$ ($1\,\%$) with $k_2 = K_2$, $k_1 = K_1$. (Black box):
$\delta= 0.1$, $K_1 = 0.001$, $K_2 = 0.1$. (Blue Dot): $\delta= 0.1$, $K_1 =
0.001$, $K_2 = 0.05$. (Green star): $\delta= 0.1$, $K_1 = 0.001$, $K_2 = 0.01$.
(Red triangle): $\delta= 0.01$, $K_1=10^{-4}$; $K_2 = 0.01$. (Dotted lines):
slopes predicted by an expansion of Eq. (\ref{eq_approx_U}) for small $\Delta
\phi_s$.
}
\end{center} 
\end{figure}

The Stern layer modifies the swimming speed for $V_0<0$ by reducing the reaction
rate 
and the potential change in the diffuse layer, appearing in Eq. (\ref{eq_U}).
Fig. \ref{fig_5} demonstrates that $\tU$ decreases for our rate constants almost
linearly with the
average Stern layer voltage drop $\langle\Delta \phi_s(\vartheta)\rangle$. The
approximation, Eq. (\ref{eq_approx_U}), does not yield satisfactory absolute
values for $\tU$ in Fig. \ref{fig_5} since we 
have here $k_1 = K_1$ and $k_2=K_2$. However, Eq. (\ref{eq_approx_U}) can be expanded
for small $\Delta
\phi_s$. It then predicts the slopes of the decrease of $\tU$ with $\Delta \phi_s$
fairly well. The inset of Fig. \ref{fig_5} demonstrates that $\Delta \phi_s$ depends non-linearly on on the
thickness of the layer $\lambda_s/\lambda$. This non-linear relationship occurs since the Stern layer voltage
drop is determined through a transcendental equation (see appendix, Eq.
(\ref{eq_determine_gamma})).
 
\begin{figure}[ht]
\begin{center}
\includegraphics[scale=0.63]{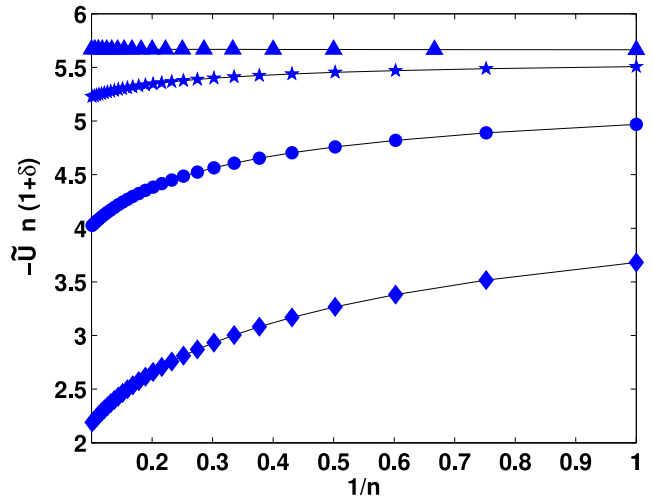}
\caption{\label{fig_6}
Deviation of $\tU$ from the standard thin diffuse layer scaling with the inverse 
ion concentration $\sim 1/(\calC + \calC \delta) $. Bulk salt concentrations are
varied as $\calC = n\times(5\times 10^{-5}\;\mathrm{mol/L})$ with $n=1\ldots10$.
Speed is multiplied with $n(1+ \delta)$ 
to remove the standard scaling. $c_{H_2O_2} = 3.7$ ($3.7 \,\%$), $K_1=
k_1=0.005$, $K_2 = k_2=0.025/n$, $\delta= 0.1/n$.
 (Triangle): $\lambda_s/\lambda = 0$.  (Star): $\lambda_s/\lambda = 0.005\,\sqrt{n}$. (Dot):
$\lambda_s/\lambda = 0.025\,\sqrt{n}$. (Diamond): $\lambda_s/\lambda = 0.1\,\sqrt{n}$.
}
\end{center} 
\end{figure}

The swimming speed scales within the thin diffuse layer model, Eq. (\ref{eq_U}),
as the inverse of the overall bulk ion concentration $1/(\calC + \calC\delta)$. 
Experimental data supports this rough scaling, but also indicates deviations from
it \cite{paxton2006catalytically}. One reason for a deviation may be the
Stern layer, which influences the reaction rate. In our simple model, the connection between the
Stern layer and the reaction is given through the appearance of $\Delta\phi_s$ in the Butler-Volmer Eq. (\ref{eq_alpha}). In a
tempting first approximation, we here keep the absolute thickness of the Stern
layer constant. The variations of the bulk salt concentration still affects the Stern layer 
voltage drop $\Delta\phi_s$ since $\lambda_s/\lambda \sim
\sqrt{\calC}$. Fig. (\ref{fig_6}) demonstrates that this suffices to cause
a reduction of the swimming speed compared with the standard scaling.  A precise
measurement of
this effect, along with a more refined model of the ion adsorption is called for
to improve the quantitative understanding here.
\section{Discussion}
Effects of salt on self-electrophoretic swimming become important when (bio-)
technological applications are envisaged since physiologically or otherwise
relevant environments are often ionic. One of these effects is the coupling
of salt ion fluxes to the reaction-driven proton fluxes. In general, salt ions can not be 
described by a constant concentration background since this approximation causes non-negligible 
errors in the swimming speed. A full mathematical description leads to a system of non-linear equations, that must
be solved numerically. We chose to focus on the thin diffuse layer approximation
rather than solving the whole system of equations. The advantages of this
approach are that the hyodrodynamics become relatively simple and that a quite
demandingly high density of grid nodes near the surface can be avoided. 
The thin diffuse layer model may describe the swimming 
well, even in absence of electrolytes. Spontaneous dissociation of $H_2O_2$ in the bulk
can be an important factor here. A reasonable pH value of 6 translates, e.g., to a double layer thickness, Eq.
(\ref{eq_debye_len}), of about $90 \, \rm{nm}$. This number is an order of
magnitude smaller than $\calR$. It justifies the approximative use of the thin
diffuse layer model for $H_2O_2$ concentrations, say, above the $2 \,\%$ range.
A contamination of the water with atmospheric carbon dioxide has also been found to
be important for electrophoresis in a salt-free environment\cite{carrique2009effects}. 
The somewhat unknown details of the reaction mechanism at the surface of the
swimmer could further contribute to making the thin diffuse layer model valid. It
is, e.g., possible that the reactions at the swimmer's surface lead, beyond our 
model, to a further accumulation of other types of ions and
radicals. 

A new feature of our model is that we use first order Butler-Volmer kinetics for
the electrocatalytic decomposition of $H_2O_2$. This reaction leads to
a linear increase of swimming speed at low  $H_2O_2$ concentrations. The transport of protons around the swimmer can 
limit the reaction rate and therefore cause a saturation of the swimming speed at high $H_2O_2$ concentrations. 
We found that an expected decrease of the pH value with an increase of the $H_2O_2$ concentration can widely suppress the leveling off of the speed 
since the bulk $H^+$ ions take part in the catalytic reduction. Any other mechanism providing an excessive amount of $H^+$ ions would equally reduce the effect 
of proton diffusion and lead to a linear speed-$[H_2O_2]$ relation. The influence of proton transport on the reaction is not in conflict with an 
additional limitation of the rates though saturation of the catalytic surface in experimental systems.

The Butler-Volmer equation includes a further 
effect, that is also related to the presence of salt: the Stern layer on the surface of the swimmer. In this
article, we have undertaken a very first step towards an understanding of its
influence on the swimming speed. The Stern layer lowers in our model the swimming speed
by modifying the potential drop across the diffuse layer and by reducing the
reaction rate. It is also responsible for a decrease of speed with 
increased salt concentration that is stronger than the inverse of the bulk ion concentration.

The swimmer's potential $V$ plays an important role for the swimming speed since
it is directly related to the potential drop across the diffuse layer (the zeta
potential). For negligible Stern layers, $V$ could possibly be determined
experimentally by measuring the additional drift of actively swimming particles
in an external field. In our model, we found $V$ to be in most cases much lower
than $ -40\, \rm{mV}$, which is about the zeta potential of a metal 
particle without the presence of $H_2O_2$. This value was used previously as a rough
estimate\cite{paxton2005motility}. Such a high potential $V$ can only be found within our model 
if the overall oxidation and reduction rates are similar (see Eq. (\ref{eq_lowest_order_V})).

In this article, we have made a few simplifying assumptions. First, we
consider a spherical swimmer with a reactivity varying like a cosine across the
surface. This geometry is chosen for its numerical robustness since no 
sharp edges and sudden changes are present. On the other hand, the 
experimentally studied swimmers are short, bimetallic rods where the reactivity varies
quite suddenly across the boundary from one metal to the other. A second 
simplificating assumption is that all solutes have similar diffusivities. We guess  
that these simplifications together introduce a deviation from the
experimental reference data by a factor of $1\,-\,5$. 

We have not modeled the adsorption of salt into the Stern layer explicitly
but fixed its width at a molecular lengthscale. The influence ion adsorption on 
electrophoretic mobility is a long standing problem. A number of different models have been suggested to
quantitatively account for different kinds of salt ions \cite{hunter1987foundations, mangelsdorf1990effects}. 
The common assumption that the Stern layer is independent of salt concentration is somewhat justified for
non-reactive, metallic interfaces\cite{grahame1954differential} but remains an assumption for
our system. Its validity may be particularly questioned once a variation of the
salt concentration changes the reaction rate and thereby the swimmer's
potential. Including an explicit adsorption mechanism is challenging. It would
be desirable to have more experimental data concerning the
interplay of salt and $H_2O_2$ decomposition at microswimmers before attempting
such a model.

It is well known that double layers can exhibit lateral ion transport if they
are highly charged \cite{bikerman1940electrokinetic}. If contributions of the
order
$\lambda \, e^{-\Delta \phi_0/2}$ are not negligible, corrections to the
$\lambda^0$ theory must incorporate electro-osmosis and electromigration in the
diffuse layer. We have neglected the
corrections due to these surface conduction phenomena in order to focus on the
important trends of our model. An investigation of the role of surface transport
for self-electrophoresis would be an important extension.
A somewhat related topic is the possible occurrence of hydrodynamic slip on the
surface of the swimmer. This slip might serve to enhance the swimming speed 
\cite{ajdari2006giant,VinogradovaPhysRevLett.107.098301}. Finally, the assumption of 
dilute solutes could be relaxed. However, non-electric interactions 
and the competition for catalytic binding sites would make an analytical 
description of solutes fairly complicated. Molecular dynamics simulations could
offer a promising alternative here.

Taken together, our model quantitatively supports an explanation of the particle swimming in terms of
self-electrophoresis \cite{paxton2006catalytically, wang2006bipolar}
and points to a number of interesting phenomena governing the details of this
intriguing mode of micro-motion.
\begin{acknowledgments}
We thank A. Sen, T.E. Mallouk and W.F. Paxton for useful comments.
\end{acknowledgments}
\appendix
\section{Details of the redox reactions}\label{sec_app_reaction}
In this appendix, we illustrate the reaction rate postulated in Sec.
\ref{sec_rates}
with concrete, but simplified prototypical reaction pathways.
We emphasize that the schemes discussed below merely present one possible way to
justify the mathematical model we discuss in the main part. Due to the general
lack of knowledge concerning the surface reactions, we can not claim that
that these pathways are ultimately the dominant ones. However, 
preliminary studies with other rate equations made us confident that many
qualitative trends presented above are quite robust.

During the reduction of $H_2O_2$, $H^+$ ions in the vicinity of the metal surface 
combine with electrons form the metal substrate $S$. A possible scheme for this process is
\begin{align}
\begin{split}
 S + H_2O_2 + H^+ + e^-&\rightarrow S(OH) + H_2O,\\
S(OH)  + H^+ + e^-&\rightarrow S + H_2O, \label{react_reduction}
\end{split}
\end{align}
which is has been used previously for a simulation of the reduction on platinum electrodes\cite{mukouyama1999mechanism}.
Additional processes, that possibly slow down the reduction, 
are neglected here. We assume that the concentration of $H_2O_2$ is low, such that
the availability of catalyst is not limiting. Applying the steady state condition,
we find that the concentration of $S(OH)$ is independent of $[H^+]$. This
implies that the effective rate equation for $H^+$ consumption is first order in
$[H^+]$ and $[H_2O_2]$. A second order reaction rate can not be excluded, but we
deem the suggested presence of a multi-step decomposition sufficient
to allow the simpler modeling of the reduction as a first order process. 

The oxidation of $H_2O_2$ at metal electrodes has been studied in great
detail by Hall et al.\cite{hall1997electrochemical, hall1998electrochemical}. It
is also found to depend on surface oxide films. A possible scheme for the reduction of the surface film is
\begin{equation}
 \begin{split}
  S(OH)_2 +  H_2O_2 \rightleftarrows S(OH)_2\cdot  H_2O_2 \rightarrow S + 2 H_2O
+ O_2.
 \end{split}
\end{equation}
Hydrogen ions are released when the surface is oxidized again 
\begin{equation}
 \begin{split}
 S +2 H_2O \rightarrow S(OH)_2 + 2 H^+ + 2 e^-.
 \end{split}
\end{equation}
Neglecting side processes, one can again assume that the reaction is first
order in $[H_2O_2]$. Four electron processes, as suggested by Wang et al.
\cite{wang2006bipolar} are, for simplicity, not considered explicitly.
\section{Nonlinear theory for $\lambda \rightarrow 0$}\label{sec_app_thin_layer}
This appendix details the non-linear mathematical theory of a thin diffuse layer
 which is used in the main part of the article. We define new variables for the
concentrations and expand them in powers of $\lambda$
\begin{align}
\begin{split}
m^{\{H,i\}} &\equiv
c^{\{H,i\},+}-c^{\{H,i\},-} \\
	       &=m^{\{H,i\}}_0 +\lambda \,
m^{\{H,i\}}_1 +\ldots,\\
M^{\{H,i\}} &\equiv
c^{\{H,i\},+}+c^{\{H,i\},-}\\
	       &=M^{\{H,i\}}_0 +\lambda \,
M^{\{H,i\}}_1 +\ldots,\\
\phi &= \phi_0 +\lambda \, \phi_1 +\ldots\,.
\label{eq_def_mMp}\\
\end{split}
\end{align}
Outer variables carry a hat ($\,\hat{\,\,}$), variables for the fields inside
the diffuse layer do not carry a hat. The potential $\phi_0$ is
written as sum of an inner and outer solution
\begin{equation}
\phi_0 = \psi_0 + \hpsi_0.
\end{equation}
The equations determining the outer variables
$\hM^{\{H,i\}}_0$, $\hm^{\{H,i\}}_0$ and $\hpsi_0$ are written in terms of
$r$ and $\vartheta$. 
The Poisson-Boltzmann equation (\ref{eq_PB}) in the outer region reads with $\hm
\equiv \hm^H +
\hm^i$ 
\begin{align}
 \nabla^2 \left( \hpsi_0 + \lambda \hpsi_1 \ldots \right) = -\frac{1}{2
\lambda^2}\left(  \hm_0 + \lambda \hm_1 + \lambda^2 \hm_2 \ldots \right),
\label{eq_outer_psi_dgl}
\end{align}
which is singular for $\lambda \rightarrow 0$ and therefore we require
$\hm_0= \hm_1 =0$. This provides the condition of charge neutrality outside the
diffuse layer. The number of lowest order outer variables is hence reduced by one
\begin{equation}
 \hm_0^H = -\hm_0^i.
\end{equation}
Note that the condition of charge neutrality does not require that the charges
of the different ions balance individually. The relation $\hm^H = \hm^i= 0$ only
holds in the bulk, far away from the particle.
Concerning the singular behavior of Eq. (\ref{eq_outer_psi_dgl}) we also mention
that a charge of (vanishing) magnitude $\sim \lambda^2 \sim 1/\calC$ results in
a non-vanishing lowest order electric field in the outer region. Therefore, the
electric potential can not be neglected in the diffusion equations, 
Eqns. (\ref{eq_diff_cp},\ref{eq_diff_cm}), outside the diffuse layer.
Eqns. (\ref{eq_diff_cp},\ref{eq_diff_cm}) read to lowest order for the outer region
\begin{align}
\nabla\cdot\left( \nabla\hm^H_0 +\hM_0^{H}\nabla \hpsi_0\right) &=0
,\label{eq_jmH}\\
\nabla\cdot\left( -\nabla\hm^H_0 +\hM_0^{i}\nabla \hpsi_0\right) &=0
,\label{eq_jmi}\\
\nabla\cdot\left( \nabla\hM_0^{H} +\hm^H_0\nabla \hpsi_0\right) &=0
,\label{eq_jMH}\\
\nabla\cdot\left( \nabla\hM_0^{i} -\hm^H_0\nabla \hpsi_0\right) &=0
,\label{eq_jMi}
\end{align}
The convection terms do not appear in lowest order because the fluid velocity
is $O(\lambda^2)$ (see below). The boundary conditions for $r \rightarrow
\infty$ are given by the bulk concentration of ions as
\begin{align}
\hM_0^H(r\rightarrow \infty,\vartheta) = 2\,\delta,\label{eq_bc_M_at_inf} \\
\hM_0^i(r\rightarrow \infty,\vartheta) = 2,\\
\hm_0^H(r\rightarrow \infty,\vartheta) = 0,\\
\hpsi_0(r\rightarrow \infty,\vartheta) = 0\label{eq_bc_Psi_at_inf}.
\end{align}
Near the surface we employ a stretched, radial coordinate $y \equiv
(r-1)/\lambda$. The diffusion equations in the inner region are expanded up to
$O(\lambda^0)$ and read
\begin{align}
0=&-\p_y\left(\p_y m^{\{H,i\}}_0 + M^{\{H,i\}}_0 \p_y \psi_0 \right),\\
0=&-\p_y\left(\p_y M^{\{H,i\}}_0 + m^{\{H,i\}}_0 \p_y \psi_0 \right).
\end{align}
The lowest order ion flux is $O(1)$. However, the fluxes pertaining to 
concentrations of $O(\lambda^0)$ vanish, since the radial derivative introduces
a 
factor of $1/\lambda$
\begin{align}
-\left(\p_y m^{\{H,i\}}_0 + M^{\{H,i\}}_0\p_y\psi_0 \right)_{y=0} = 0,\\
-\left(\p_y M^{\{H,i\}}_0 + m^{\{H,i\}}_0\p_y\psi_0 \right)_{y=0} = 0.
\end{align}
The leading order result for the variables $m_0^{\{H,i\}}(y,\vartheta)$,
$M_0^{\{H,i\}}(y,\vartheta)$ in the diffuse layer are then given by
\begin{align}
m_0^{\{H,i\}}(y,\vartheta) &=
a^{\{H,i\}}(\vartheta)\,e^{-\psi_0(y,\vartheta)}-b^{\{H,i\}}(\vartheta)\,e^{
\psi_0(y,\vartheta)}, \label{eq_inner_m}\\
M_0^{\{H,i\}}(y,\vartheta) &=
a^{\{H,i\}}(\vartheta)\,e^{-\psi_0(y,\vartheta)}+b^{\{H,i\}}(\vartheta)\,e^{
\psi_0(y,\vartheta)} \label{eq_inner_M},
\end{align}
with yet unknown constants.
Matching between the diffuse layer and the outer region  as $\lim_{y\rightarrow
\infty} \{m^{\{H,i\}}_0,M^{\{H,i\}}_0\} = \lim_{r\rightarrow 1}
\{\hm^{\{H,i\}}_0,\hM^{\{H,i\}}_0\} $ yields 
\begin{align}
a^{\{H,i\}}(\vartheta)-b^{\{H,i\}}(\vartheta) = \hm_0^{\{H,i\}}(1,\vartheta),\\
a^{\{H,i\}}(\vartheta)+b^{\{H,i\}}(\vartheta) = \hM_0^{\{H,i\}}(1,\vartheta).
\end{align}
The number of constants $a^{\{H,i\}},\; b^{\{H,i\}}$ in our solutions is larger
than the number of equations. One needs to consider the matching of the
$O(\lambda)$ solutions in order to obtain the full lowest order solution.
The result of this further matching process are conservation equations for the
radial solute fluxes, which we do not derive here for brevity
\begin{align}
 \left(-\p_r \hM_0^H(r,\vartheta)-\hm^H_0(r,\vartheta)\p_r
\hpsi_0(r,\vartheta)\right)_{r=1} &= \alpha_0(\vartheta),
\label{eq_bc_MH_at_1}\\
  \left(-\p_r \hM_0^{i}(r,\vartheta)+\hm^H_0(r,\vartheta)\p_r
\hpsi_0(r,\vartheta)
\right)_{r=1} &= 0, \label{eq_bc_Mi_at_1}\\
 \left(-\p_r \hm^H_0(r,\vartheta)-\hM_0^{H}(r,\vartheta)\p_r
\hpsi_0(r,\vartheta)
\right)_{r=1} &= \alpha_0(\vartheta), \label{eq_bc_mH_at_1}\\
 \left( \p_r \hm^H_0(r,\vartheta)-\hM_0^{i}(r,\vartheta)\p_r
\hpsi_0(r,\vartheta)
\right)_{r=1} &= 0. \label{eq_bc_mi_at_1}
\end{align}
In the inner region, the Poisson-Boltzmann equation is non-singular. It reads 
up to $O(\lambda^0)$
\begin{align}
\p^2_y \psi_0(y,\vartheta) = -\frac{m_0}{2} = B(\vartheta)
\sinh{\psi_0(y,\vartheta)},
\end{align}
with 
\begin{equation}
B(\vartheta)=
{\left(\hM_0^{H}(1,\vartheta)+\hM_0^{i}(1,\vartheta)\right)}/2.
\end{equation}
This equation, determining the non-equilibrium concentration near the surface,
has the same form as the corresponding equilibrium equation. The result for the
potential
in the diffuse layer is 
\begin{equation}
\begin{split}
\phi_0(y,\vartheta) =&  \psi_0(y,\vartheta) + \hpsi_0(1,\vartheta)= \\
&2 \ln\left[\frac{1+\gamma(\vartheta) e^{-\sqrt{B(\vartheta)}
y}}{1-\gamma(\vartheta) e^{-\sqrt{B(\vartheta)} y}}\right] +
\hpsi_0(1,\vartheta).\label{eq_inner_psi0_appendix} 
\end{split}
\end{equation}
The value of the potential at the outer periphery of the diffuse layer
$\hpsi_0(1,\vartheta)$ is determined through the solution of Eqns.
(\ref{eq_jmH}-\ref{eq_jMi}).
Given $\hpsi_0(1,\vartheta)$, the overall change in potential over the diffuse
layer $\Delta \phi_0(\vartheta)$
is according to Eq. (\ref{eq_phi_stern_condition}) 
\begin{equation}
  \Delta \phi_0(\vartheta) \equiv \psi_0(0,\vartheta) = V_0 - \Delta
\phi_s(\vartheta) - \hpsi_0(1,\vartheta) \label{eq_stern_bc}.
\end{equation}
The voltage drop over the Stern layer is calculated through Eq.
(\ref{eq_determine_phi_s}), which 
is in terms of inner variables
\begin{equation}
\Delta\phi_s(\vartheta) = -\frac{\lambda_s}{\lambda} \frac{\p
\psi_0(y,\vartheta)}{\p y}|_{y =0} =\frac{\lambda_s}{\lambda} \sqrt{B(\vartheta)}
\frac{\gamma(\vartheta)}{1-\gamma(\vartheta)^2}. \label{eq_phi_s_formula}
\end{equation}
Insertion of Eq. (\ref{eq_phi_s_formula}) into Eq. (\ref{eq_stern_bc})
yields a transcendental equation for the function $\gamma(\vartheta)$ 
\begin{equation}
V_0=\frac{\lambda_s}{\lambda} \sqrt{B(\vartheta)}
\frac{\gamma(\vartheta)}{1-\gamma(\vartheta)^2} + 2
\ln\left[\frac{1+\gamma(\vartheta)}{1-\gamma(\vartheta)}\right] +
\hpsi_0(1,\vartheta) \label{eq_determine_gamma}
\end{equation}
which now allows the full determination of the potential in the diffuse layer.
In order to calculate the lowest order swimming speed in $\hbe_z$-direction from
the above results we expand Eq.(\ref{eq_teubner_U}) for small
$\lambda$.  The lowest order result is
\begin{align}
\begin{split}
U_0 = g\lambda^2\int_0^{\infty}\int_0^{\pi}\frac{m_0(y,\vartheta)}{2} [y^2
\p_y \phi_0(y,\vartheta)\cos\vartheta \\
-y\p_{\vartheta} \phi_0(y,\vartheta) \sin\vartheta] 
\sin\vartheta \,\rmd \vartheta \rmd y. \label{eq_u_teubner_thin_dbllayer} 
\end{split}
\end{align}
On inserting Eqns. (\ref{eq_inner_m},\ref{eq_inner_M},\ref{eq_inner_psi0})
including the matched constants and using partial integration we find Eq.
(\ref{eq_U}) for the swimming speed.
An explicit calculation of $U$ requires the knowledge of the
outer fields at $r=1$. Therefore, the pertaining differential Eqns.
(\ref{eq_jmH}-\ref{eq_jMi}) need to
be solved with the boundary conditions Eqns.
(\ref{eq_bc_M_at_inf}-\ref{eq_bc_Psi_at_inf},
\ref{eq_bc_MH_at_1}-\ref{eq_bc_mi_at_1}). Also, the potential of the swimmer,
$V_0$, depends through the charge conservation Eq.
(\ref{eq_charge_conservation})
on $\alpha$ and must be determined simultaneously with the concentration fields.
 The whole task can only be done
numerically. To this end, we implement an efficient pseudospectral
method\cite{boyd2001chebyshev} where the discretized system of differential
equations is solved with the Newton-Raphson method. In radial direction, we use
a Chebychev grid with 30
nodes. The grid in angular direction, for $\vartheta=0\ldots\pi$, is
uniformly spaced with 40 nodes.  
\section{Analytical approximation for $U$} \label{sec_app_approx}
In this appendix we provide the details for the approximate swimming speed,
which is given in Sec. \ref{sec_approx_U}. Due to radial symmetry and the
boundary conditions at infinity no ion fluxes
arise in the lowest order where $k_1\approx0, \, k_2 \approx 0$. Although being a non-equilibrium steady
state, the concentrations of ions follow an equilibrium distribution throughout
the whole system. Solutions for the outer field variables, far away from the
swimmer, are easily determined from Eqns. (\ref{eq_jmH}-\ref{eq_jMi}) with 
$\hpsi_{0,0} = 0$ to be $\hM_{0,0}^H = 2\delta$, $\hM_{0,0}^i = 2$, $\hm_{0,0}^H
= \hm_{0,0}^i=0$. The indices denote the $\lambda^0$ 
and $k_1^0,k_2^0$ order. The inner fields of lowest order obey the differential
equations
\begin{align}
0=&-\left(\p_y m^{H,i}_{0,0} + M^{H,i}_{0,0} \p_y \phi_{0,0} \right),\\
0=&-\left(\p_y M^{H,i}_{0,0} + m^{H,i}_{0,0} \p_y \phi_{0,0} \right),
\end{align}
and the solutions are given by Eqns. (\ref{eq_inner_m},\ref{eq_inner_M}) with
$a^H_{0,0}=b^H_{0,0}=\delta$, and $a^i_{0,0}=b^i_{0,0}=1$. The potential
$\phi_{0,0}$ is 
\begin{equation}
\begin{split}
\phi_{0,0}(y) = 2 \ln\left[\frac{1+\gamma_0 e^{-\sqrt{B_0} y}}{1-\gamma_0
e^{-\sqrt{B_0} y}}\right],\label{eq_inner_psi0_approx} 
\end{split}
\end{equation}
with $B_{0} = \delta +1$ and $\gamma_0 = \tanh(\phi_{0,0}(0)/4)$. 

For the outer variables of first order in $k_1,\,k_2$ we demand $\hm_{0,1}^i =-\hm_{0,1}^H$, as in App. \ref{sec_app_thin_layer}. The
field equations read 
\begin{align}
\nabla^2 \hM^H_{0,1} =\nabla^2 \hM^i_{0,1} &= 0,\\
\nabla^2 \hm^H_{0,1} =\nabla^2\hpsi_{0,1} &=0.
\end{align}
The corresponding boundary conditions are
\begin{align}
\begin{split}
 -\p_r\hM_{0,1}^H|_{r=1} &= \alpha_{0,1},\\
-\p_r\hM_{0,1}^i|_{r=1} &= 0,\\
-\left(\p_r\hm_{0,1}^H -2\delta \p_r\hpsi_{0,1}\right)|_{r=1} &= \alpha_{0,1},\\
\left(\p_r\hm_{0,1}^H -2 \p_r\hpsi_{0,1}\right)|_{r=1} &= 0\\
 \hM_{0,1}^H|_{r=\infty} =\hM_{0,1}^i|_{r=\infty}=\hm_{0,1}^H|_{r=\infty} &= 0,\\
\hpsi|_{r=\infty} &= 0.
\end{split}
\end{align}
Taken together, these equations yield
\begin{align}
\hM^i_{0,1} &= 0,\\
\hM^H_{0,1} &= \sum_{n=0}^{\infty} A_n\,P_n(\cos\vartheta)\,r^{-n-1},\\
\hm^{H}_{0,1} &=\hM^H_{0,1}/(\delta +1),\\
\hpsi_{0,1} &= \hM^H_{0,1}/(2 \delta +2),
\end{align}
where the the coefficients of the expansion in Legendre polynomials
$P_n(\cos\vartheta)$ are 
calculated from
\begin{equation}
-\p_r \hM^H_{0,1}|_{r=1} = \sum_{n=0}^{\infty} \left(n+1\right)A_n P_n(\cos \vartheta) =
\alpha_{0,1}. \label{eq_first_order_alpha_bc}
\end{equation}
For the corresponding inner variables of first order in $k_1,\,k_2$ we have the
differential equations
\begin{align}
0=&-\left(\p_y m^{H,i}_{0,1} + M^{H,i}_{0,1} \p_y \phi_{0,0} + M^{H,i}_{0,0} \p_y
\phi_{0,1}\right),\\
0=&-\left(\p_y M^{H,i}_{0,1} + m^{H,i}_{0,1} \p_y \phi_{0,0} + m^{H,i}_{0,0} \p_y
\phi_{0,1}\right),
\end{align}
which result in 
\begin{align}
m_{0,1}^{H,i} = a_{0,1}^{H,i}e^{-\phi_{0,0}}- b_{0,1}^{H,i} e^{\phi_{0,0}} -
\phi_{0,1} M_{0,0}^{H,i},\\
M_{0,1}^{H,i} = a_{0,1}^{H,i}e^{-\phi_{0,0}}+ b_{0,1}^{H,i} e^{\phi_{0,0}} -
\phi_{0,1} m_{0,0}^{H,i}.
\end{align}
Matching now as $\lim_{y\rightarrow \infty} M^{H,i}_{0,1} =
\hM^{H,i}_{0,1}|_{r=1}$, $\lim_{y\rightarrow \infty} m^H_{0,1} =
\hm^H_{0,1}|_{r=1}$, $\lim_{y\rightarrow \infty} \phi_{0,1} =
\hpsi_{0,1}|_{r=1}$ we find $b_{0,1}^{i}=-a_{0,1}^{i}=0$, $b_{0,1}^{H}=0$ and 
\begin{equation}
a_{0,1}^{H}=  \sum_{n=0}^{\infty} A_n P_n(\cos \vartheta). \label{eq_a_A_matching}
\end{equation}
This yields for the $H^+$ concentration at the outer edge of the compact layer
\begin{equation}
c^{H+}_{0,1} = \frac{M_{0,1}^{H} + m_{0,1}^{H}}{2} = \left(a_{0,1}^{H} - \delta
\phi_{0,1}\right)e^{-\phi_{0,0}}.
\end{equation}
Having obtained the concentration of cations at the surface of the swimmer, we
are now in the position to calculate the reaction rate $\alpha_{0,1}$ in first
order of $k_1,\, k_2$. We insert $c^{H+}_{0} \approx \delta e^{-\phi_{0,0}} +
c^{H+}_{0,1}$ in Eq. (\ref{eq_alpha}) and employ Eq.
(\ref{eq_lowest_order_alpha}) for the lowest order reaction rate. The boundary
condition for the potential Eq. (\ref{eq_lowest_order_V}) in order  $k_1^0, \,
k_2^0$ is also employed along with the corresponding also in first order Stern
layer boundary condition
\begin{align}
 V_{0,1} = \phi_{0,1}(0,\vartheta)  + \Delta \phi_{s,1}(\vartheta)
\end{align} 
to yield
\begin{align}
\alpha_{0,1} = e^{\frac{\Delta \phi_{s,0}}{2}} \left( - \frac{a^H_{0,1}
K_2}{\delta} + K_2 V_{0,1} - \cos\vartheta\left(k_2 + k_1
\frac{K_2}{K_1}\right)\right).
\end{align}
This expression along with Eqns. (\ref{eq_first_order_alpha_bc},
\ref{eq_a_A_matching}) allows to determine $a^H_{0,1}$. Since the swimmer can
not emit a net flux of cations, we demand that the radially symmetric part of
$\alpha_{0,1}$ vanishes. Therefore we find with Eqns.
(\ref{eq_first_order_alpha_bc}, \ref{eq_a_A_matching}), $A_0=0$, $V_{0,1} =0$,
and 
\begin{align}
a^H_{0,1} = -\delta \frac{   k_2 + k_1 \frac{K_2}{K_1} }{2 \delta \,
e^{-\frac{\Delta \phi_{s,0}}{2}} + K_2} \cos\vartheta.\label{eq_a_first_order}
\end{align}
In order to calculate the swimming speed we first consider the body force in the
diffuse layer
\begin{align}
\begin{split}
 m_0 \nabla \phi_0 \approx &m_{0,0} \nabla \phi_{0,0}+(m_{0,1}^H+m_{0,1}^i)\, \hbe_r \,\p_r \phi_{0,0} \\
	      &+(1+\delta)(e^{-\phi_{0,0}}-e^{\phi_{0,0}}) \nabla \phi_{0,1}.
\end{split}
\end{align}
The first, lowest order term is radially symmetric and therefore irrelevant for the swimming.
On inserting the expressions for $m_{0,1}$, we find 
\begin{align}
\begin{split}
 m_0 \nabla \phi \approx &m_{0,0} \nabla \phi_{0,0} + a_{0,1}^H e^{-\phi_{0,0}} \,\hbe_r \, \p_r \phi_{0,0} \\
&+(1+\delta)\nabla\left((e^{-\phi_{0,0}}-e^{\phi_{0,0}}) \phi_{0,1}\right).
\label{eq_bf_first_order}
\end{split}
\end{align}
Since the last term in the body force is a gradient it only modifies the
hydrodynamic pressure and
does cause motion of the swimmer in the incompressible limit. Therefore, the
first order potential $\phi_{0,1}$ is irrelevant for the swimming speed.
Inserting Eq. (\ref{eq_bf_first_order}) with Eq. (\ref{eq_a_first_order}) into
Eq. (\ref{eq_u_teubner_thin_dbllayer}), we find the formula for the swimming
speed up to first order in the asymmetry of surface reactivity given in the main
text.

%
\end{document}